# Anomaly Detection in XML-Structured SOAP Messages Using Tree-Based Association Rule Mining

Reyhaneh Ghassem Esfahani, Mohammad Abadollahi Azgomi[*] and Reza Fathi

Trustworthy Computing Laboratory, School of Computer Engineering,
Iran University of Science and Technology, Tehran, Iran
E-mail: r_isfehani@comp.iust.ac.ir, azgomi@iust.ac.ir and fathi_reza@comp.iust.ac.ir

## Abstract

Web services are software systems designed for supporting interoperable dynamic cross-enterprise interactions. The result of attacks to Web services can be catastrophic and causing the disclosure of enterprises' confidential data. As new approaches of attacking arise every day, anomaly detection systems seem to be invaluable tools in this context. The aim of this work has been to target the attacks that reside in the Web service layer and the extensible markup language (XML)-structured simple object access protocol (SOAP) messages. After studying the shortcomings of the existing solutions, a new approach for detecting anomalies in Web services is outlined. More specifically, the proposed technique illustrates how to identify anomalies by employing mining methods on XML-structured SOAP messages. This technique also takes the advantages of tree-based association rule mining to extract knowledge in the training phase, which is used in the test phase to detect anomalies. In addition, this novel composition of techniques brings nearly low false alarm rate while maintaining the detection rate reasonably high, which is shown by a case study.

## 1. Introduction

In service-oriented architecture (SOA), each part of the system is implemented in the form of a service or a disjoint unit such that there is little dependency on underlying technologies used in implementing them. Although using SOA brings less maintenance and reusability costs, exposing it in open environments, such as the Internet, can cause instabilities in using these services because of failure, security attacks and so on.

Web services, as a realization of SOA, are commonly used not only in the Internet, but also in intranets and inter-organization communications. They usually provide access to vital systems, such as enterprise resource planning, to which a successful attack may cause heavy damages. Therefore, it is important to consider appropriate mechanisms to protect these services from both accidental failures and security attacks.

---

[*] Address for correspondence: School of Computer Engineering, Iran University of Science and Technology, Hengam St., Resalat Sq., Tehran, Iran, Postal Code: 16846-13114, Fax: +98-21-73021480, E-mail: azgomi@iust.ac.ir.

Intrusion detection systems (IDSs) are proposed for the detection of attacks and intrusions to computer networks. Two major approaches of analyzing data in intrusion detection systems are *misuse detection* and *anomaly detection*. In the former approach, attack detection is done on the basis of a set of predefined attack patterns or signatures stored in a database. In the latter, the IDS defines a normal profile of the system under supervision and tries to identify any deviations from it. As a result, this approach is capable of identifying new unseen attacks as well.

A special kind of IDSs works in Web service layer and is called *Web services intrusion detection system* (WS-IDS). WS-IDSs are used to protect Web services from malicious attacks. To design new enhanced WS-IDSs, proposing new techniques for anomaly detection in Web services is a key requirement.

In this paper, after an investigation on shortcomings of the existing solutions, a new approach for detecting anomalies in Web services is outlined. More specifically, the proposed technique illustrates how to identify anomalies by employing mining methods on extensible markup language (XML)-structured *simple object access protocol* (SOAP) messages. This technique also takes the advantages of *tree-based association rule mining* to extract knowledge in the training phase, which is used in the test phase to detect anomalies. In addition, this novel composition of techniques brings nearly low false alarm rate while maintaining the detection rate reasonably high. This technique is implemented and experimented in a case study. The experiment results are also presented in the paper.

The rest of this paper is organized as follows. First, we have a brief review on Web services terminology in Section 2. After a short look at related works in this area in Section 3, we highlight our motivations in Section 4. In Section 5, we explain our proposed technique. Then, the proposed technique is evaluated through a case study in Section 6. Finally, Section 7 concludes the paper and outlines future works.

## 2. Background

In this section, we briefly define some concepts and terms, which are used throughout the paper.

### 2.1. Web services

According to the World Wide Web consortium (W3C), a Web service is a software system designed to support interoperable machine-to-machine interaction over a network. It has an interface described in a machine-processable format especially in web service definition language (WSDL). Other systems interact with the Web service in a manner prescribed by its description using SOAP messages, typically conveyed using the hypertext transport protocol (HTTP) with an XML serialization in conjunction with other Web-related standards [1].

### 2.2. Simple object access protocol

Simple object access protocol (SOAP) is an Internet-based platform-independent protocol that provides a way to communicate between applications. Previous application communication protocols, such as remote procedure call (RPC), were raised some security and compatibility issues. Therefore, firewalls and proxy servers, usually block their traffic. SOAP messages, on the other hand, are sent over HTTP, which is a transfer protocol supported by all Internet browsers and servers. Simplicity and extensibility are other features of this XML-based protocol [2].

A SOAP message is an ordinary XML document containing the following elements:

- *Envelope*: the root element that identifies the XML document as a SOAP message.
- *Header*: an optional element that contains the information like authentication or payment.
- *Body*: a required element that contains *call* and *response* information.
- *Fault*: an optional element, containing errors and status information.

Each SOAP request is embedded in data section of an application layer protocol to be transferred over the Internet. SOAP requests are usually transferred trough *GET* and *POST* methods of HTTP protocol.

## 2.3.    Web services and SOAP security

Because firewalls and gateways see SOAP messages as a normal HTTP traffic, it is possible for these messages to get around firewalls. Therefore, any possible malicious contents in SOAP messages will remain undetected and this fact can jeopardize the Web services security.

In [3], attacks on Web services are classified into three main classes: (1) infrastructure attacks, which are attacks related to web servers where the Web services reside and also the attacks related to the transport protocol used for exchanging the Web services' request, (2) Web services attacks, that are native to the actual technology fueling Web services, such as WSDL scanning, and (3) XML content attacks, which can be any type of XML-based, content-driven threat that employ the tactic of embedding malicious content with a legitimate XML document.

In [4], security problems of XML Web services are investigated and the result of their studies is presented as eight categories of possible attacks on Web services. The categories are as follows [4]:

- *Identity attacks*: dictionary, IP spoofing, message eavesdropping and data tampering attacks.
- *Session attacks*: replay and man-in-the-middle attacks.
- *Parsing attacks*: recursive payloads, oversize payloads and schema poisoning attacks.
- *Buffer overflow* attacks.
- *Code attacks*: SQL injection and XPath injection attacks.
- *XML denial of service (XDoS) attacks*.
- *WSDL attacks*: WSDL scanning and parameter tampering attacks.
- *Internal attacks*: These attacks perform within the firewalls by an employee or former employee that usually have access to confidential information and are familiar with internal system of the organization.

According to [5], there are three logical layers for security implementation in SOAP messages:

1. *Transport layer*, which include protocols for exchanging SOAP messages. This approach has the limitation of point-to-point interaction, but what we need is message-level security to support the multi-application interaction. Another drawback is that no uniform standard exists to carry security context from transport layer to application layer.
2. *Application layer*, which refers to service providers and service consumers. Implementing custom security mechanisms in applications, besides being error prone, can cause integration problems when two services are to be composed.
3. *SOAP layer*, which is the engines that help exposing business logic in applications as SOAP-based services. SOAP layer seems to be the only alternative where the security model can be built.

OASIS, a nonprofit consortium in charge of developing open standards for information society, proposed WS-Security which is a set of enhancements to SOAP messaging to provide message integrity and confidentiality [6]. Their specified mechanisms could accommodate a wide variety of security models and encryption technologies. Although WS-Security has improved the security of SOAP messages, it was the base of new threats such as denial of service (DoS) [4].

As stated above, standards are just suggestions to meliorate the situation and other remedies to secure vital Web services must be sought.

## 2.4.    Intrusion detection systems

Any set of actions that attempt to compromise the integrity, confidentiality, or availability (CIA) of a resource is termed as an intrusion [7]. An intrusion detection system is a system that tries to identify intrusions. Based on the analysis models, intrusion detection systems fall in two main distinguished categories: misuse detection and anomaly detection. Misuse detection systems are designed to identify previously known attacks by using a set of predefined patterns. The main shortcomings of such systems are the need to update attack signatures database and also their inability in detecting novel unseen attacks. Anomaly detection systems, on the other hand, are capable of detecting new unseen attacks. Anomaly detection is based on the belief that any malicious activity will cause a change in the normal pattern of resource usage. Therefore, to set up an anomaly detection system, a comprehensive normal pattern of the system behavior must be at hand, and providing such patterns may not always be an easy job.

## 2.5.    Tree-based association rule mining

The concept of *association rules* was popularized particularly due to [8]. Association rules describes co-occurrence of data items in a dataset and are usually represented in the form of $X \Rightarrow Y, X \Rightarrow Y$ where $X$ and $Y$ are two arbitrary different data items. This co-occurrence is a pattern and also interesting if it validates a hypothesis that the user sought to confirm. An interesting pattern represents knowledge and there are two measures of rule support and confidence to measure a rule's interestingness [9].

Rule support corresponds to the frequency of $X \cup Y$ while $X \cap Y = \emptyset$ in the dataset which is $p(X \cup Y)$. Notice that the notation $P(X \cup Y)$ indicates the probability that a dataset contains the union of set $X$ and set $Y$. Rule confidence corresponds to the conditional probability of finding $Y$ having found $X$ and is given by $p(Y|X) = \frac{sup(X \cup Y)}{sup(X)}$ where $sup(x)$ is the number of the occurrence of $X$ in the dataset [10].

The element-only XML Information Set (Infoset) content model [11] is considered to get the normal pattern using association rules. Infoset allows an XML nonterminal tag include only other elements and/or attributes and terminal elements is confined to the text. Further, some features of the Infoset, which are irrelevant to our work, such as namespaces, the ordering label, the referencing formalism through ID-IDREF attributes, URIs and links, are not considered. An XML document is represented by a labeled tree <$N$, $E$, $r$> where $N$ is the set of tree nodes, $E$ is the set of tree' edges and $r$ is the root node. An example of an XML document and its tree-based representation and some sub-trees are shown in the Figure 1 [12].

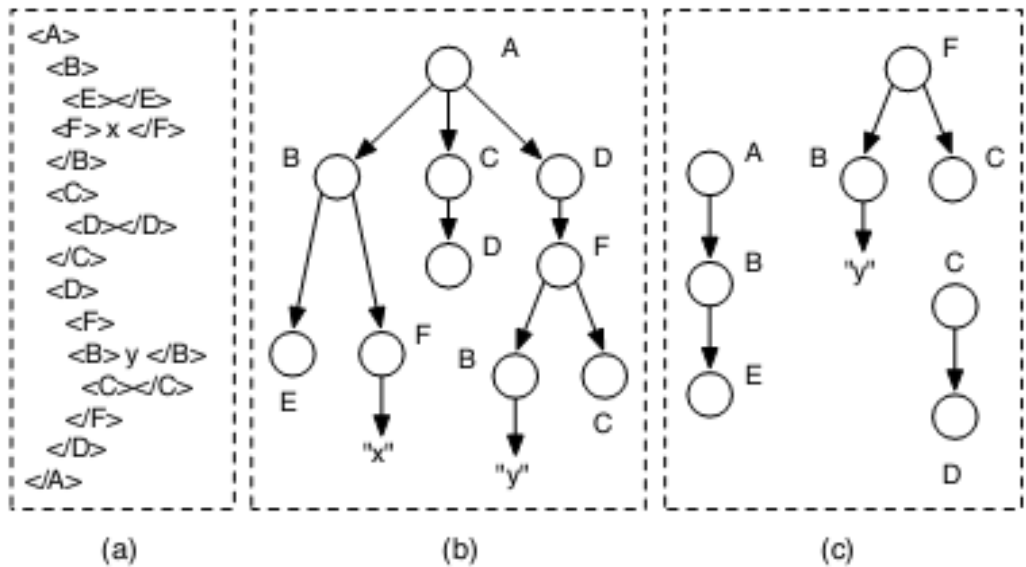

**Figure 1.** (a) an XML document example, (b) its tree-based representation, and (c) three induced sub-trees **[12]**

## 3. Related works

Anomaly detection is the process of finding patterns in data that do not conform to expected behavior and is an important problem that has been the topic of many surveys and articles as well as books. In [13], an application-level gateway, called Checkway, is suggested to prevent DoS attacks by grammatical validation on SOAP messages before forwarding them to the server. The proposed method limits the overall size of SOAP messages, or in a more complicated way, limits the size and the number of XML elements in messages. This approach may dismiss valid messages too. In [14], an intrusion detection and prevention framework is proposed that uses agents as sensors to gather suspected items. To analyze these items, clustering, association rule-based and sequential rule-based techniques together with fuzzy logic are used. In another work [15], a distributed hierarchical multi-agent system is proposed that takes the advantage of decision-tree and neural network classification mechanisms in two phases. The main advantage of this approach is said to be the adaptation ability to the changes that occur in the patterns of attacks. There are also researches on XML firewalls, as in [16], where a firewall for Web services was built. Three filtering policies of message size filtering, syntax parsing, and XML schema validation were implemented in the firewall. In [17], a simple anomaly-based Web Application Firewall (WAF) is presented. With the aid of an XML file that specifies the Web application normal behavior, statistically characterized, the WAF decides whether the incoming requests are malicious or not. Their experiments with artificially generated traffic showed that when the normal behavior is closely characterized, near perfect detection rate is reached. In [18], architecture of a Web service firewall is proposed that supports authentication and authorization mechanisms. It also provides prevention of SOAP-based attacks.

Open Web Application Security Project (OWASP) [19], is a not-for-profit worldwide open-source project that tries to provide unbiased, practical, cost-effective information about application security. OWASP foundation does not endorse or recommend commercial products or services, allowing the community to remain vendor agonistic with the collective wisdom of the best minds in application security worldwide. In [20], after explaining the security requirements of Web services, it is discussed how different standardization bodies address these requirements. This work also investigates the issues related to the use of the standards and open research issues in the area of access control for Web services and digital identity management techniques.

Many intrusion detection and prevention systems (IDS/IPS) are designed and implemented to secure computer networks. A survey and taxonomy of different techniques used in such systems is given in [21]. Several other works exist in the literature, which review intrusion detection techniques introduced for different layers of networks.

In [22], the architecture of a Web service intrusion detection system is proposed, which encompasses both anomaly detection and misuse detection techniques. Hidden Markov models are used to detect anomalies. However, the proposed concept should be validated in a real world environment. In [23] and [24], a multi-layer architecture for intrusion-tolerant web services is presented that uses single version software fault tolerance concepts in the case of malicious failures. The proposed architecture is evaluated using a Petri net model.

There are also vast researches on employing data mining techniques in intrusion detection systems, a survey of which can be found in [25]. Here, we take a short review at the knowledge extraction techniques from XML documents.

In [26], the S-MAS architecture is introduced, which is an adaptive approach for dealing with DoS attacks in Web Service environments. S-MAS is a distributed hierarchical multi-agent architecture that implements a classification mechanism in two phases. The main benefits of the approach are the distributed capabilities of the multi-agent systems and the self-adaption ability to the changes that occur in the patterns of attack. A prototype of the architecture was developed and the results obtained are presented in this study.

The approach proposed in [27] makes use of two security testing techniques, namely penetration testing and fault injection, in order to emulate XSS attack against Web services. This technology, combined with WS-Security and security tokens, can identify the sender and guarantee the legitimate access control to the SOAP messages exchanged. They have used the vulnerability scanner *soapUI* that is one of the most recognized tools of penetration testing. They have also introduced WSInject as a new fault injection tool, which introduces faults or errors on Web services to analyze the behavior in an environment not robust.

Many attempts have been done on automatically saving XML documents in the form of relational object-oriented data [28] or native XML databases [29]. As a result, many query languages were emerged. But the problem is that users of the query languages should exactly know what type of information they are looking for besides the obstacle that the input and output of these query languages are limited. Using data mining enables users to find out unknown and hidden facts from raw data. XML documents' main characteristics are their structure and semantic, which necessitate new aspects of analysis in data mining process. Mining XML documents is different from mining structured documents.

In [30], it is described that XML mining process is consist of three main phases: preprocessing, pattern extraction and post processing. XML data preprocessing infers related structure and content from available sources and then mining techniques are used to identify patterns. The output of this process is often a tree or a graph representation that shows the document structure or schema. To extract patterns, classification, clustering and association rule mining algorithms can be employed on the preprocessed data and then in post processing phase, extracted patterns are validated and interpreted. In [31], the key problem of discovering frequent patterns is discussed and the way of adjusting an efficient, scalable, and parallelizable tree pattern mining algorithm to reduce its memory requirements compensating time complexity is described.

In recent years, association rule mining has gained lots of attentions in XML data analysis such as in [28], [32], [33], [34], [35], [36].

In [33], an extension of XQuery for mining association rules, called XMINE RULE was proposed. This operator declaratively specified complex mining tasks on relational data. In other words, XML data had to be translated into a relational form. Also, the user had to specify the head and body structure of the rule to be mined which is not a logical requirement. Another shortcoming of the XMINE RULE is that the extracted rules have one specific root, and when this root is determined, only its descendants will be analyzed.

In [34], the authors have tried to extract association rules, using only a specially devised hierarchical data structure, called HoPS, without multiple XML data scans or any prior knowledge about the XML data. These rules were called XML association rules and represented in the form of $X \Rightarrow Y$, where $X$ and $Y$ are parts of the XML document. $X$ and $Y$ also had to be disjoint. It should be noted that $X$ and $Y$ are embedded sub-trees of the XML document and such sub-trees only maintain the ancestor-descendant relationship, not the parent-child one. Therefore, the extracted rules do not always show the real structure of data.

In [35], tree-based association rules (TAR) were mined from XML documents. A TAR represents knowledge in the forms of $S_B \Rightarrow S_H$, where $S_B$ is the body tree and $S_H$ is the head tree of the rule. The rule states that if the tree $S_B$ appears in an XML document $D$, it is likely that the bigger or more specific tree, $S_H$, also appears. In [34], the mined knowledge is approximate and intentional and is used to provide quick, approximate answers to queries. This knowledge also provides information about structural regularities that can be used as data guides for document querying. The mined rules hold the parent-child relationship and are stored in XML format so that they can be queried later on.

# 4. Motivations and aims

In service-oriented architecture (SOA), the traffic often has to pass through firewalls. Firewalls are means of packet filtering that operate in third and fourth layers of the OSI reference model, where TCP/IP headers are analyzed. This type of firewalls are suitable to prevent DoS attacks such as "TCP/SYN flood" and "ping of death", which exploit TCP/IP protocols.

Application level gateways are designed to analyze application protocols above the fourth layer. These gateways can identify simple application protocols such as HTTP and can protect services against malformed HTTP requests.

Obviously, the above mentioned solutions cannot fully protect Web services. The packet filtering and application gateway approaches only investigate HTTP and TCP/IP headers, while Web services use XML-structured SOAP messages in the application layer headers and so TCP/IP packet filtering systems cannot identify XML-based attacks. On the other hand, the inherit security weakness of SOAP, Universal Description, Discovery and Integration (UDDI) and WSDL, which are the basic building blocks of Web services, can be targets of attacks [37].

To improve the situation, a set of standard extensions of SOAP, called WS-Security [6] was proposed. WS-Security tries to implement message integrity and confidentiality with the use of XML encryption and XML signature. Such standards may fulfill parts of the Web services security needs, but unfortunately, due to some complexities, they may cause new vulnerabilities. In this situation, there is an urgent need for intrusion detection systems that can detect attacks to XML messages. As mentioned before, everyday intruders seek new methods to attack Web services. Therefore, to have a safer environment, Web services must be secure against new unseen attacks too.

Having the existing methods, techniques and tools to secure Web servers and databases, the aim of this work has been to target the attacks that reside in the Web service layer and the XML-based SOAP messages, and does not consider attacks related to the underlying infrastructure and technologies of Web services.

We intend to use the anomaly detection approach to analyze the SOAP traffic so that we can find any possible attacks. The main requirement of an anomaly detection system is a precise and complete description of the normal behavior of the system under supervision. This means that to detect anomalies from SOAP messages, a normal profile of these messages must be at hand. There are many approaches to define normal profiles in anomaly detection systems, and one of the most common ones is data mining [38]. The point is that SOAP messages are in the form of XML documents and general data mining techniques cannot directly operate on XML-based data.

XML is a simple, flexible text format derived from standard generalized markup language (SGML), and is designed as a means of exchanging wide variety of data on the Web and elsewhere. XML tags define the structure and semantics of information within the document and that is the reason why XML documents are semi-structured and self-describing. The more widespread XML documents becomes, the more difficult becomes extracting knowledge from them.

To find a solution, the rest of this paper investigates the existing methods of using data mining on XML files and the goal is to extract knowledge from SOAP messages and then build a normal profile for a Web service anomaly detection system.

## 5. The proposed method

In this section, we introduce a new method for the detection of anomalies in XML-structured SOAP messages, which uses tree-based association rule mining. To this end, first we present some definitions, assumptions and concepts. Then, the details of the proposed method, including the required algorithms and processes will be introduced.

### 5.1. Definitions and assumptions

Web services are placed in the demilitarized zone of network firewall. Nowadays, using tools such as XML firewalls or Web service firewalls are very common. Web service firewall ensures that the traffic is secure and then sends it to the Web server of the Web service. Figure 2 shows the place where Web services intrusion detection system (WS-IDS) stands among other elements of network.

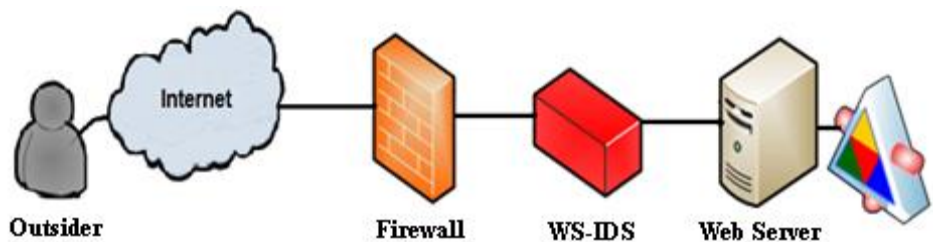

**Figure 2.** The placement of WS-IDS in the network

At an abstract level, an anomaly is defined as a pattern that does not conform to the expected normal behavior [39]. Therefore a simple approach to detect anomaly is defining the area that shows the normal behavior and all the data that do not belong to this area is called anomaly. Based on the labeled data available, anomaly detection methods can be divided into three categories [39]:

- Supervised anomaly detection: techniques that operate in this mode assume that the training dataset has labeled instances both for normal and anomalous classes.
- Semi-supervised anomaly detection: techniques of this group, which are more widely used, need labeled instances only for the normal class.
- Unsupervised anomaly detection: techniques that fall into this category do not require training data at all and thus are most widely applicable. In this category, it is assumed that normal instances are far more frequent than anomalous ones.

In our case, we choose semi-supervised approach, which requires only the normal instances, because defining all anomalous instances is impossible as every day a new way of attacking Web services arises.

The assumptions of the proposed technique are as follows:

1. In our technique, like other anomaly detection techniques, it is assumed that behavioral patterns of malicious attacks are different from normal transaction patterns.
2. Authentication of service consumers is done in previous steps, probably by a firewall or any other mechanisms and the proposed technique will not take any action to authenticate and authorize users. Therefore, any intruder who has bypassed this step cannot be detected as long as he/she has normal behavior.
3. The components of the anomaly detection system are highly reliable and they are not compromised by intruders.
4. None of the WS-* standards are used to protect SOAP messages.

## 5.2.    Knowledge extraction from XML documents

To perform semi-supervised anomaly detection on SOAP messages of a Web service, we have to provide a normal profile of correct SOAP requests and responses, for training the anomaly detection system. As explained before, SOAP messages are XML-based. The body part of the SOAP message will be used.

Therefore, we must choose a data mining technique to extract patterns from these XML-based data. We exploit association rule mining algorithm to find the rules that govern the normal instances in the training dataset and in this way the structure of the normal profile is formed.

After studying the existing methods, we concluded that tree-based association rule mining best matches our needs for anomaly detection in SOAP messages. One prominent reason is the scale of the data, which is high and the other one is huge number of candidates to be considered for finding out the rules. In the following subsection, we present the details of implementing the proposed technique.

## 5.3.    Details of the proposed method

The aim has been to take the advantage of tree-based association rule (TAR) mining technique and apply it to the SOAP message instances in the training phase. The obtained information about structural regularities can be the base of the normal profile.

According to [34], extracting TARs has two main steps: (1) mining frequent sub-trees from the SOAP messages of the training set and (2) extracting rules from the obtained sub-trees of the previous step.

Taking the algorithm's efficiency into account, the authors of Dryadeparent in [40] showed that their algorithm is the fastest tree mining algorithm available and *CMTreeMiner* [41] stands in the second position. A specification of Dryadeparent is that it holds ancestor-descendant relationship, but what we need here is the parent-child relation, which is supported by *CMTreeMiner*. Therefore, our selected algorithm for mining sub-trees will be *CMTreeMiner*. Figure 3 depicts our algorithm of mining TARs from SOAP messages, which is based on TAR extraction algorithm in [34] and also inspired by [30] XML mining framework. This algorithm receives the SOAP messages training set and identifies their frequent sub-trees by *Get-Frequent-Trees* procedure and then in the next part, association rules that are hidden in these frequent structures are mined by the *Mine-Association-Rules* procedure. The *Mine-Association-Rules* procedure calls *Extract-Rules* procedure in order to extract a rule from a subtree which has a support value higher than a given confidence by the min-confidence parameter. The sup procedure calculates the rule's support needed to be calculated inside the *Extract-Rules* procedure. This obtained knowledge can be used in the test phase to identify potentially malicious SOAP requests. In addition, we use the regular expression (RE) of the frequent structures' content - to define the ranges of values and strings instead of each possible value - to detect anomaly request from normal ones in the next phase as shown in Figure 5. Figures 4 and 5 show a general overview of how the proposed technique works in the training and test phases.

```
 Get-Frequent-Trees (SOAP-Messages)
INPUT: normal traffic of SOAP messages
OUTPUT: frequent subtrees
1:  subtree set F_S, DFCFs = ∅
2:  preprocess the gathered messages to remove unimportant parts
3:  for all soap ∈ SOAP_Messages do
4:       DFCFs = The depth-first traverse of the preprocessed messages
5:  end for
6:  F_S = CMTreeMiner(DFCFs)
7:  return F_S

 Mine-Association-Rules (F_S, minconf)
INPUT: frequent subtrees and minimum confidence of rules to be mined
OUTPUT: tree-based association rules
1:  set rules mined_rules = ∅
2:  for all s ∈ F_S do
3:       tempSet = Extract_Rules(s, minconf)
4:       mined_rules = mined_rules ∪ tempSet
5:  end for
6:  return mined_rules

Extract - Rules(s, minconf)
INPUT: a tree of frequent subtrees set
OUTPUT: possible new rule or else ∅
1:  rule set mined_rules = ∅
2:  for all c_s subtree of s do
3:       conf = sup(s)/sup(c_s)
4:       if conf ≥ minconf then
5:           newRule = < c_s, s, conf, sup(s)>
6:            mined_rules = mined_rules ∪ newRule
7:       end if
8:  end for
9:  return  mined_rule
```

**Figure 3.** The proposed TAR extraction algorithm from SOAP messages

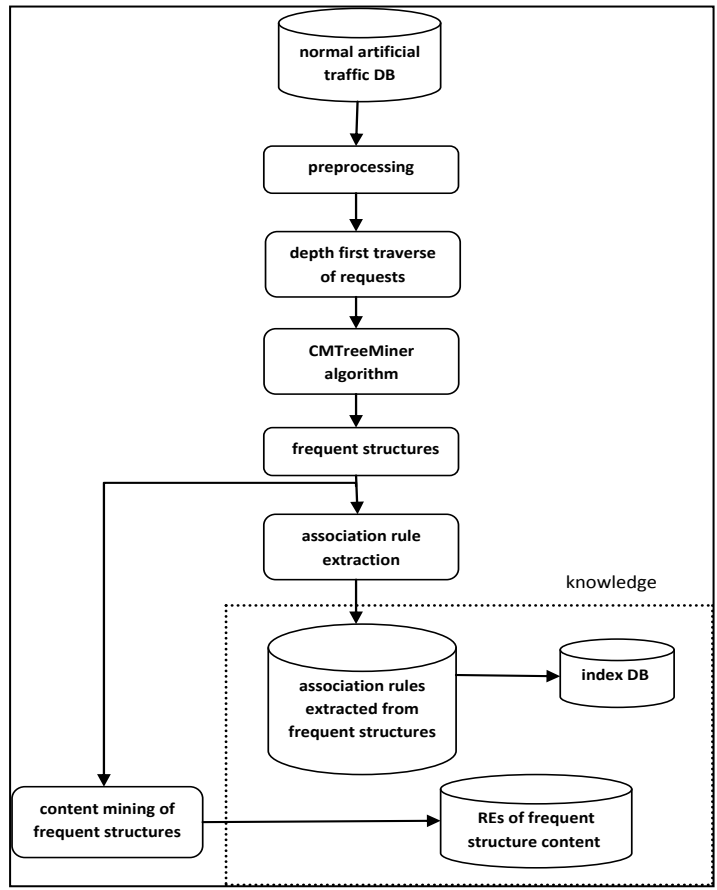

**Figure 4.** Steps of the training phase

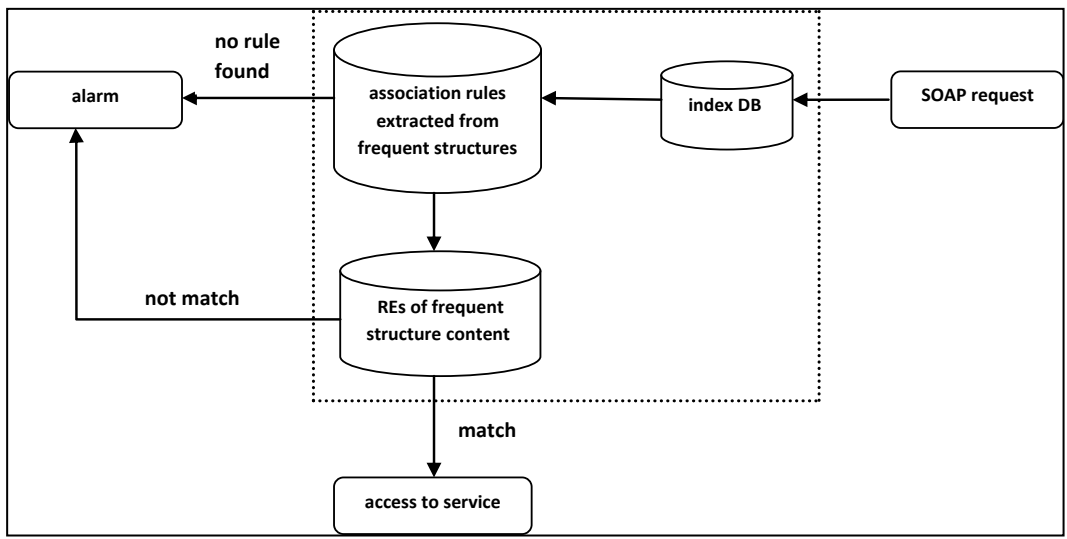

**Figure 5.** Steps of the test phase

## 6. Case study

To the best of our knowledge, since there is no standard method or tool available to measure the efficiency of WS-IDS by the time of writing this paper, we have implemented the proposed method and test it on an experimental Web service.

An XML structured sample dataset available on the internet is the Wikipedia's article categories graph [42]. In Wikipedia each concept is defined by an article and these articles are categorized into different classes that best describe their content, so that there can be a quick access to the articles. In addition to articles, each category in Wikipedia is at least the member of a more general category. In this way, all articles and categories in Wikipedia form a big graph, which is called "Category Graph" and its root is a category named "Category:Contents". Figure 6 shows parts of this category graph.

We have defined a Web service on this dataset that has a Web method called "GetWikiSubCategory". This method requires a category name as input and the output is the available subcategories for that category to the depth of 3.

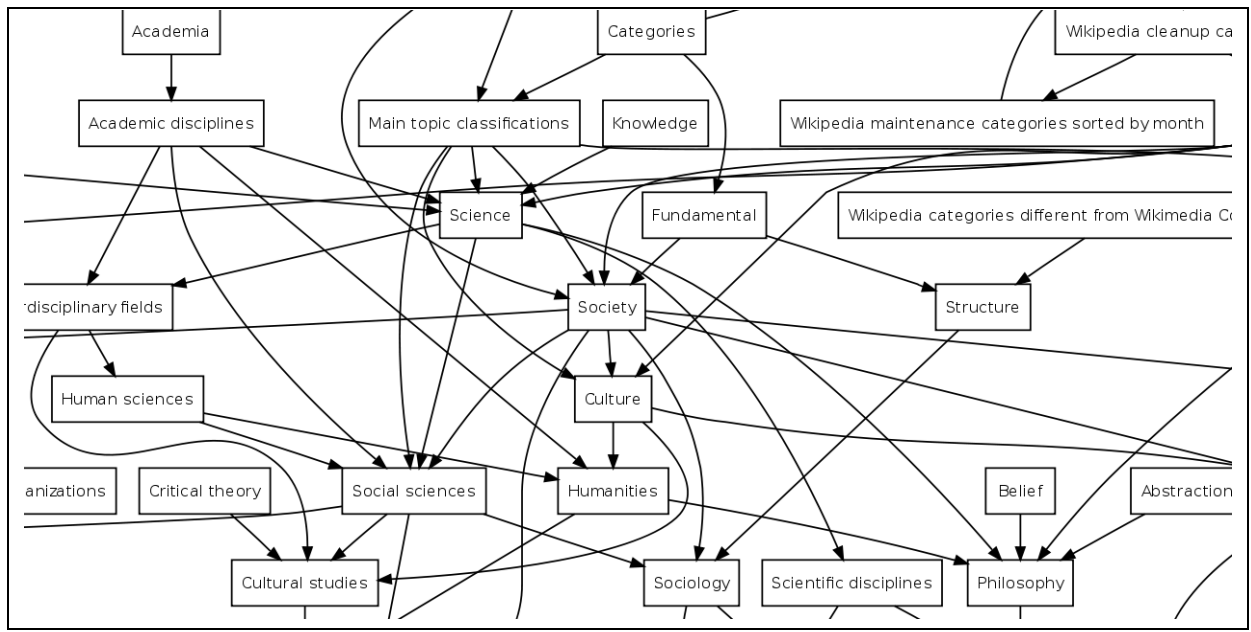

**Figure 6.** Parts of the Wikipedia's category graph [43]

It is worth mentioning that there are two types of categories: one is content categories that help readers to find articles and project categories that are used by editors or automated tools, based on the features of the current state of articles. Figure 7 shows a small part of the content category, which we have selected for our experiments. The other one is structure category that tries to extract knowledge from the structure graph of the category.

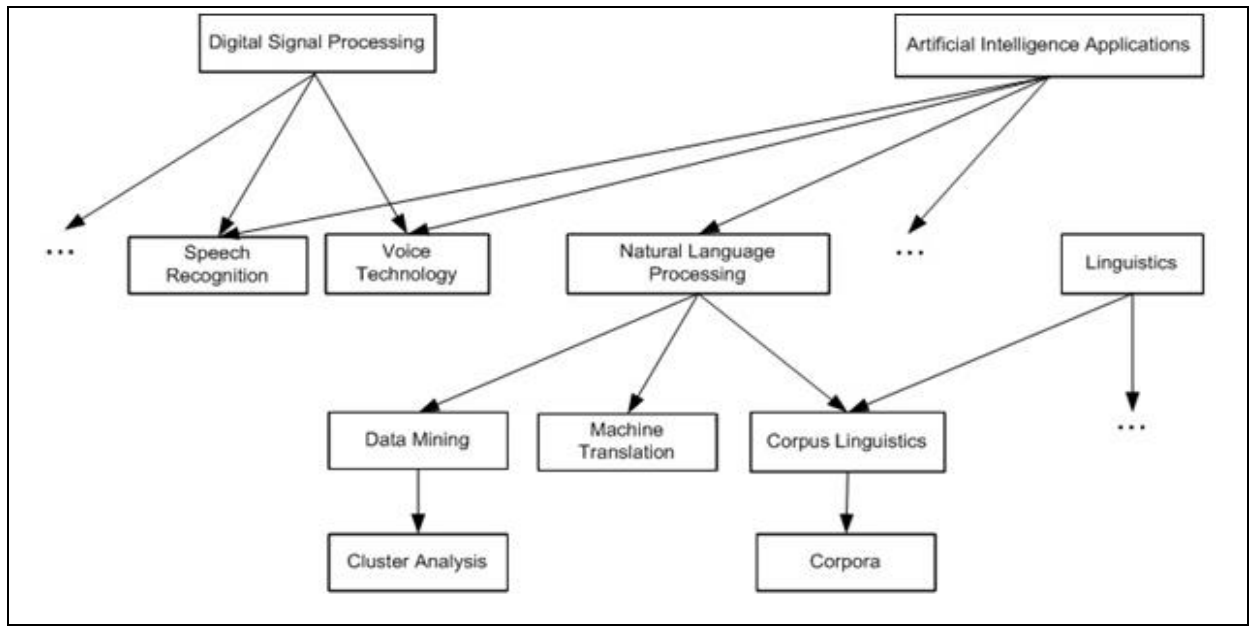

**Figure 7.** Parts of the categories used in this study

In the next step, we have generated the normal SOAP requests and responses by sending a category name to the Web service and getting its subcategories to the depth of 3.

Then, at the training phase, in six stages we trained the system by 50, 100, 150, 200, 250, and at last by 300 normal SOAP requests and responses. Then, the depth-first traversals of the generated SOAP messages are given to *CMTreeMiner* algorithm. After obtaining the frequent sub-trees, the *Mine-Association-Rule* algorithm will generate the rules. Figure 8 describes the overall structure of the XML file containing the generated rules. For instance, *rule0* states that if the tree shown in the *<CS>* tag appears, it is probable that the tree in *<S>* tag appears too. The confidence and support of each rule are also provided in the following tags. Figures 9(a) and 9(b) depict two samples of the mined association rules.

```
<rules>
        <rule0>
                <Cs> rule0's body </Cs>
                <S> rule0's header </S>
                <confidence> rule0's confidence </confidence>
                <support> rule0's support </support>
        </rule0>
        <rule1>
                ...
        </rule1>
        ...
</rules>
```

**Figure 8.** Overall view of the file containing the extracted rules

The XML document of rules is then stored in the "*association rules extracted from frequent structures*" database as showed in Figure 4. Then, we made an index of these rules and stored it in "*index*" database, so that the rules can be retrieved fast. Also, the regular expressions of the frequent structures' content are placed in the "*REs of frequent structures content*" database.

```
<rule0>

        <Cs>
                <Natural_language_processing>
                        <Machine_translation></Machine_translation>
                        <Data_mining></Data_mining>
                        <Corpus_linguistics></Corpus_linguistics>
                </Natural_language_processing>
        </Cs>
        <S>

                <soap:Envelope>
                        <soap:Body>
                                <GetWikiSubCategory>
                                        <Natural_language_processing>
                                                <Machine_translation></Machine_translation>
                                                <Data_mining></Data_mining>
                                                <Corpus_linguistics></Corpus_linguistics>
                                        </Natural_language_processing>
                                </GetWikiSubCategory>
                        </soap:Body>
                </soap:Envelope>
        </S>
        <confidence>0.5</confidence>
        <support>2</support>
</rule0>
```

(a)

```
<rule1>
        <Cs>
                <Natural_language_processing>
                        <Machine_translation></Machine_translation>
                        <Data_mining></Data_mining>
                        <Corpus_linguistics></Corpus_linguistics>
                </Natural_language_processing>
        </Cs>
        <S>
                <soap:Envelope>
                        <soap:Body>
                                <GetWikiSubCategory>
                                        <Artificial_intelligence_application>
                                                <Voice_technology>
                                                        <Speech_recognition></Speech_recognition>
                                                        <Speech_synthesis></Speech_synthesis>
                                                <Voice_technology>
                                                <Speech_recognition></Speech_recognition>
                                                <Natural_language_processing>
                                                        <Machine_translation></Machine_translation>
                                                        <Data_mining></Data_mining>
                                                        <Corpus_linguistics></Corpus_linguistics>
                                                </Natural_language_processing>
                                        </Artificial_intelligence_application>
                                </GetWikiSubCategory>
                        </soap:Body>
                </soap:Envelope>
        </S>
        <confidence>0.5</confidence>
        <support>2</support>
</rule1>
```

(b)

**Figure 9.** Two samples of the extracted tree-based association rules ((a) and (b))

We also need some anomalous SOAP messages for the test phase. The number of anomalous requests is considered constantly 50 throughout all experiments, so that the fluctuations of the detection rate can be observed as we increase the number of training data. In the anomalous requests generation process, it is assumed that the intruder does not have any information about the structure and content of the database. Therefore, we used either one of the following approaches to generate the anomalous requests:

1. Embedding the pattern of a known attack such as parser attacks (coercive parsing, oversize payloads and recursive payloads) and other attacks like SQL/XML injection, in the service request file.
2. Embedding anomalous patterns that do not match to any known attacks.

Then, in the test phase, as depicted in Figure 5, the anomalous requests are sent to the Web service. Afterward, for each node of the request, the "*index*" database is searched to find the index of all rules that contains that node. After retrieving the rules from the database of association rules, it is investigated that whether the structure of the received request comply with any of them or not. If not, it is probable that an attack, such as a parsing attack, is occurring and consequently an alarm is raised. Otherwise, the content of the received request is compared with the regular expressions available in the database of regular expressions. Again, if no match is found, there is a probability of attack, such as an injection attack and so an alarm will be raised; otherwise, if a match is found, the request is allowed to have access to the service.

## 7. Evaluation results

To measure the detection rate and false alarm rate more precisely, we have repeated each experiment two times, with two different compositions of training data and considered the average of these two experiments as the detection rate and false alarm rate of that run.

The ratio of detected attacks to all attacks is called detection rate (Equation 1). Also, the ratio of normal requests that are mistakenly considered as anomalous to all normal requests is called false alarm rate (Equation 2). The detection rate and false alarm rate of the proposed method in different runs of our case study is shown in Table 1. The abbreviation *TP* means detecting a true attack as an attack and *FP* means detecting a normal data mistakenly as an attack. And *TN* means detecting a normal data as a normal data and *FN* means detecting an attack wrongly a normal data.

$$DetectionRate = \frac{\text{TP}}{\text{TP+FN}} * 100\% \tag{1}$$

$$FalseAlarmRate = \frac{\text{FP}}{\text{FP+TN}} * 100\% \tag{2}$$

**Table 1.** Detection rate and false alarm rate of the proposed technique

| No. of training patterns | No. of extracted rules | No. of anomalous patterns | Detection rate | False alarm rate |
|---|---|---|---|---|
| 50 | 23 | 50 | 100% | 6% |
| 100 | 39 | 50 | 100% | 2% |
| 150 | 49 | 50 | 100% | 1% |
| 200 | 54 | 50 | 99.5% | 0.5% |
| 250 | 55 | 50 | 99.2% | 0.008% |
| 300 | 55 | 50 | 99.1% | ~ 0 |

Figure 10, horizontal axis indicates the number of data and the vertical one shows the percentage, shows the detection rate and false alarm rate of the proposed technique in different number of training instances. In primary stages, the number of training data is low and as a result, higher false alarm rate is observed. But, as the number of training instances increases, the method learns wider range of normal instances and so the false alarm rate decreases. The detection rate is almost constant during the experiments, but when the false alarm rate is getting near to zero, the detection rate falls a bit below 1, which is still acceptable. In our evaluation test, after doing it on 300 normal data, we got near zero false alarm and 99.1 percent detection rate which is an acceptable precise rate of detection.

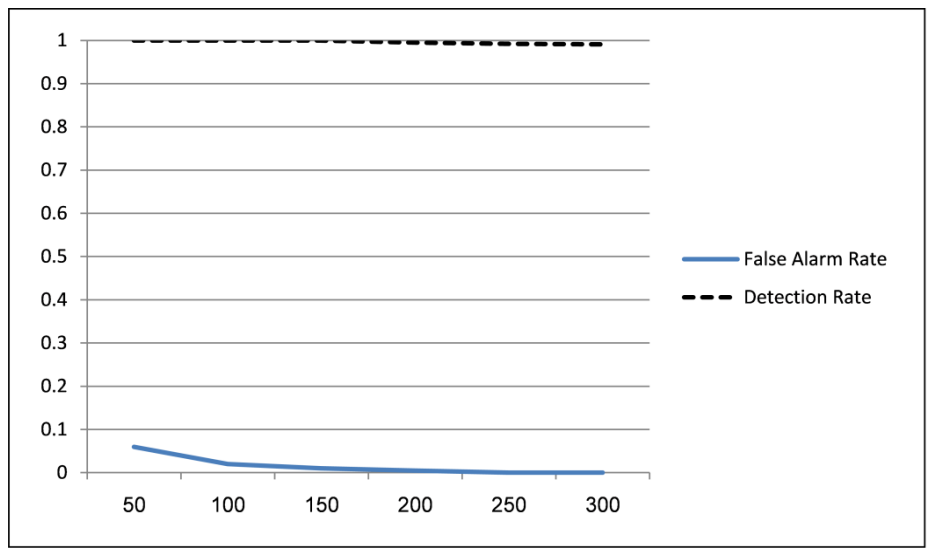

**Figure 10.** The detection rate and false alarm rate of the proposed technique

## 8. Conclusions

In this paper, we introduced a new technique for intrusion detection in Web services. The technique takes the advantages of XML mining and tree-based association rule to detect both known and new unseen attacks. The normal behavior of the Web service is learned by the system in the training phase and in the test phase this knowledge is used to detect probable attacks.

If the system is trained with less normal instances, more requests will be rejected and as a result the detection rate is almost 1, but the false alarm rate is high as well. On the other hand, when the system is learned with more normal patterns, the false alarm rate mitigates fast and the detection rate remains acceptably high.

As a future work, one can prepare a real or synthesized standard dataset, which contains different attack patterns to evaluate the effectiveness of Web services intrusion detection system. Such standard dataset can be used as a basis to compare different Web services intrusion detection system. Another idea is to store the requests that the system detects as anomalous and do further investigations on them to have insights of new attack patterns. The other work can be extending the tree-based association rule mining algorithm such that the content of frequent trees would turn to regular expressions automatically.

# References

[1]   WWW Group, "Web Services Architecture," [Online]. Available: http://www.w3.org/TR/ws-arch/#whatis. [Accessed: March 2013].

[2]   "SOAP Tutorial," January 2011. [Online]. Available: http://www.w3schools.com/soap.

[3]   A. Manzanares, "Professional Pen Testing for Web Applications - Programmer to Programmer Series (P2P)," *The Computer Journal,* vol. 50, no. 4, pp. 500-500, 2007.

[4]   E. Moradian and A. Hkansson, "Possible attacks on XML web services," *International Journal of Computer Science and Network Security (IJCSNS),* vol. 6, no. 1B, pp. 154-170, 2006.

[5]   K. Ramarao and C. Prasad, SOA Security, Manning Publications Co, 2008.

[6]   K. Lawrence, C. Kaler, A. Nadalin, R. Monzillo and P. Hallam-Baker, "Web services security: SOAP message security 1.1 (WS-security 2004)," *OASIS, OASIS Standard,* 2006.

[7]   R. Heady, G. Luger, A. Maccabe and M. Servilla, The architecture of a network-level intrusion detection system, Technical Report CS90-20, Department of Computer Science, University of New Mexico, 1990.

[8]   R. Agrawal, T. Imieli ski and A. Swami, "Mining association rules between sets of items in large databases," in *Proceedings of the International Conference on Foundations of Data Organization and Algorithms*, Chicago, 1993.

[9]   J. Han and M. Kamber, *Data mining: concepts and techniques*, 3$^{rd}$ ed., Morgan Kaufmann, 2011.

[10]  P.-N. Tan, M. Steinbach, V. Kumar, *Introduction to Data Mining*, Addison-Wesley, 2005.

[11]  W. W. W. Consortium, "XML Information Set," Second Edition, 2004. [Online]. Available: http://www.w3.org/TR/xml-infoset/.

[12]  M. Mazuran, E. Quintarelli and L. Tanca, Mining tree-based association rules from XML documents, Technical Report, Politecnico di Milano, 2009.

[13]  N. Gruschka and N. Luttenberger, "Protecting web services from dos attacks by SOAP message validation," *Security and Privacy in Dynamic Environments,* pp. 171-182, 2006.

[14]  C. Yee, W. Shin and G. Rao, "An adaptive intrusion detection and prevention (ID/IP) framework for web services," in *Proceedings of the International Conference on Convergence Information Technology*, pp. 528-534, 2007.

[15]  C. Pinzn, J. De Paz, J. Bajo and J. Corchado, "An adaptive multi-agent solution to detect dos

attack in SOAP messages," in *Proceedings of the International Conference on Computational Intelligence in Security for Information Systems,* pp. 77-84, 2009.

[16] Y. Loh, W. Yau, C. Wong and W. Ho, "Design and implementation of an XML firewall," in *Proceedings of the International Conference on Computational Intelligence and Security,* vol. 2, pp. 1147-1150, 2006.

[17] C. Torrano-Gimenez, A. Perez-Villegas and G. Alvarez, "A self-learning anomaly-based web application firewall," *Proceedings of the International Conference on Computational Intelligence in Security for Information Systems,* pp. 85-92, 2009.

[18] Z. Aliannezhadi and M. Abdollahi Azgomi, "Modeling and analysis of a Web service firewall using coloured Petri nets," in *Proceedings of the IEEE Asia-Pacific Services Computing Conference (APSCC'08),* Jiaosi, Yilan, Taiwan, 2008.

[19] Open Web Application Security Project (OWASP), [Online]. Available: http://www.owasp.org/index.php/Main_Page. [Accessed: April 2013].

[20] L. Martino and E. Bertino, "Security for web services: Standards and research issues," *International Journal of Web Services Research (IJWSR),* vol. 6, no. 4, pp. 48-74, 2009.

[21] A. Lazarevic, V. Kumar, J. Srivastava, "Intrusion detection: A survey," in *Managing Cyber Threats,* Vol. 5, Kumar, V. and Srivastava, J. and Lazarevic, A. (eds.), Springer, pp. 19-78, 2005.

[22] M. S. A. Najjar and M. Abdollahi Azgomi, "A distributed multi-approach intrusion detection system for web services," in *Proceedings of the 3rd International Conference on Security of Information and Networks*, ACM, pp. 238-244, 2010.

[23] Z. Aghajani and M. Abdollahi Azgomi, "A multi-layer architecture for intrusion-tolerant Web services," *International Journal of u-and e-Service, Science and Technology,* vol. 1, no. 1, pp. 73-80, 2008.

[24] Z. Aghajani and M. Abdollahi Azgomi, "Security evaluation of an intrusion tolerant Web service architecture using stochastic activity networks," in *Proceedings of the 3rd International Conference on Information Security and Assurance (ISA'09)*, Korea University, Seoul, Korea, 2009.

[25] C. Tsai, Y. Hsu, C. Lin and W. Lin, "Intrusion detection by machine learning: A review," *Expert Systems with Applications,* vol. 36, no. 10, pp. 11994-12000, 2010.

[26] C.I. Pinzón, J., Bajo, J. F. De Paz and J. M. Corchado, "S-MAS: An adaptive hierarchical distributed multi-agent architecture for blocking malicious SOAP messages within Web Services environments", *Expert Systems with Applications*, vol. 38, no. 5, pp. 5486-5499, 2011.

[27] M. I. P. Salas E. Martins, "Security testing methodology for vulnerabilities detection of XSS in Web services and WS-security", *Electronic Notes in Theoretical Computer Science*, vol. 302, pp. 133–154, 2014.

[28] S. Abiteboul, P. Buneman and D. Suciu, *Data on the Web: from relations to semistructured data and XML*, Morga Kaufmann Publishers, 2000.

[29] E. Pardede, J. Rahayu and D. Taniar, "Object-relational complex structures for XML storage," *Information and Software Technology,* vol. 48, no. 6, pp. 370-384, 2006.

[30] R. Nayak, R. Witt and A. Tonev, "Data mining and XML documents," in *Proceedings of the International Conference on Internet Computing*, June 24-27, 2002.

[31] A. Jiménez, F. Berzal and J.-C. Cubero, "Mining frequent patterns from XML data: Efficient

algorithms and design trade-offs", *Expert Systems with Applications*, vol. 39, no. 1, pp. 1134-1140, 2012.

[32] C. Combi, B. Oliboni and R. Rossato, "Querying XML documents by using association rules," in *Proceedings of the Sixteenth International on Workshop Database and Expert Systems Applications,* pp. 1020-1024, 2005.

[33] D. Braga, A. Campi, S. Ceri, M. Klemettinen and P. Lanzi, "Discovering interesting information in XML data with association rules," in *Proceedings of the 2003 ACM symposium on Applied computing*, pp. 450-454, 2003.

[34] J. Paik, H. Youn and U. Kim, "A new method for mining association rules from a collection of XML documents," in *Proceedings of the International Conference on Computational Science and Its Applications* pp. 1-7, 2007.

[35] K. Wang and H. Liu, "Discovering structural association of semistructured data," *IEEE Transactions on Knowledge and Data Engineering,* vol. 12, no. 3, pp. 353-371, 2000.

[36] I. Suganya, N. Velmurugan and P. Ganeshkumar, "XML query-answering support system using association mining technique", in *Proceedings the 2013 IEEE Conference on Information and Communication Technologies (ICT)*, pp. 1259-1262, 2013.

[37] C. Yee, W. Shin and G. Rao, "An adaptive intrusion detection and prevention (ID/IP) framework for Web services," in *Proceedings of the International Conference on Convergence Information Technology*, pp. 528-534, 2007.

[38] P. Garcia-Teodoro, J. Diaz-Verdejo, G. Mac-Fernndez and E. Vzquez, "Anomaly-based network intrusion detection: Techniques, systems and challenges," *Computers & Security,* vol. 28, no. 1, pp. 18-28, 2009.

[39] V. Chandola, A. Banerjee and V. Kumar, "Anomaly detection: A survey," *ACM Computing Surveys (CSUR),* vol. 41, no. 3, p. 15, 2009.

[40] A. Termier, M. Rousset, M. Sebag, K. Ohara, T. Washio and H. Motoda, "Dryadeparent, an efficient and robust closed attribute tree mining algorithm," *IEEE Transactions on Knowledge and Data Engineering,* vol. 20, no. 3, pp. 300-320, 2008.

[41] Y. Chi, Y. Yang, Y. Xia and R. Muntz, "Cmtreeminer: Mining both closed and maximal frequent subtrees," *Advances in Knowledge Discovery and Data Mining,* pp. 63-73, 2004.

[42] T. Zesch and I. Gurevych, "Analysis of the Wikipedia category graph for NLP applications," in *Proceedings of the TextGraphs-2 Workshop (NAACL-HLT'07)*, pp. 1-8, 2007.

[43] "Wikipedia:Categorization," [Online]. Available: http://en.wikipedia.org/wiki/Wikipedia:Categorization. [Accessed: April 2013].